\documentclass[12pt]{iopart}
\usepackage{iopams}
\usepackage{epsf}
\usepackage{euscript}
\usepackage{graphicx}
\usepackage{color}

\renewcommand{\epsilon}{\varepsilon}

\begin{document}

\title[Irreversible quantum graphs]%
{Irreversible quantum graphs}
\author{Uzy Smilansky}
\address{ Department of Physics of Complex Systems}
\address{The Weizmann Institute of Science, 76100 Rehovot, Israel}
\address{\vspace*{\baselineskip} E-mail: \texttt{Uzy.Smilansky@weizmann.ac.il}}
\begin{abstract}
Irreversibility is introduced to quantum graphs by coupling the
graphs to a bath of harmonic oscillators. The interaction which is
linear in the harmonic oscillator amplitudes is localized at the
vertices. It is shown that for sufficiently strong coupling, the
spectrum of the system admits a new continuum mode which exists
even if the graph is compact, and  a {\it single} harmonic
oscillator is coupled to it. This mechanism is shown to imply that
the quantum dynamics is irreversible. Moreover, it demonstrates
the surprising result that irreversibility can be introduced by a
``bath" which consists of a {\it single} harmonic oscillator.
\end{abstract}
\section{Introduction}
\label{sec:Introduction} Quantum graphs emulate many properties of
quantum chaotic systems, and their use is now widely spread in the
physics and mathematics literature  \cite{KS97, KS99,car99,
KoSch99, KS00, Kuchment02, KS03}. In many potential applications
the physical systems are irreversible, and it is desirable to
include this feature to quantum graphs, and thus enlarge their
range of applicability.

    The standard way to introduce irreversibility  to a quantum system is by
coupling it to a ``bath" which consists of a set of ``irrelevant"
degrees of freedom. The bath dynamics is governed, however, by a
well defined quantum hamiltonian \cite {Pauli,vanHove, Feynman,
Fescbach}. The enlarged system, composed of the original system
and the bath, is a proper quantum system, with hamiltonian:
\begin{equation}
\label{eq:hamiltonian}
H= H^{system} + H^{bath} +W
\end{equation}
$H^{system}$ and $H^{bath}$ are operators in the ``system" and
``bath" Hilbert spaces, respectively, and $W$ is the coupling. In
the physics literature, irreversibility is generally associated
with the bath having a  continuous spectrum.  The physical
justification is based on the observation that the Poincar\'e
recurrence  time is infinite, or stated differently, the mean
probability to return back to the initial state vanishes, which is
a prerequisite for irreversibility. This argument can also be
formulated mathematically. To simplify the presentation, assume
that $H^{system}$ has a discrete spectrum, while the bath spectrum
is continuous, with a spectral measure ${\rm d}\mu({\cal E})$ and
$\delta$ normalized eigenfunctions $|\chi_{\cal E}\rangle $.
Denoting by  $\Psi_{\cal P}$ the projection of an eigenstate of
  $H$ on the system subspace,  The Schr\"odinger equation can be
reduced to
  \begin{equation}
\label{eq:effective}
E \Psi_{\cal P} = H^{system} \Psi_{\cal P} +\lim_{\eta \searrow 0}
\int {\rmd}\mu({\cal E}) W_{\cal P}({\cal E}) \frac{1}{E-{\cal
E}+i\eta} W^{\dagger}_{\cal P}({\cal E}) \Psi_{\cal P}    \ ,
\end{equation}
where $ W_{\cal P}({\cal E})$ is the projection of $W|\chi_{\cal E}\rangle$ on the system
subspace. Using the identity
\begin{equation}
 \lim_{\eta \searrow 0}\frac{1}{x+i\eta} =-i \pi
\delta(x) +{\cal P}\left ( \frac{1}{x}\right ),
\label{eq:deltaid}
\end{equation}
one traditionally associates irreversibility  with the operator
\begin{equation}
\label{eq:dissipative}
 \Gamma(E) = -i \pi  \int {\rmd}\mu({\cal
E}) \delta(E-{\cal E}) W_{\cal P}({\cal E}) W^{\dagger}_{\cal
P}({\cal E})
  \ .
\end{equation}
$\Gamma(E) $ is an anti hermitian operator, which introduces decay
to the quantum time evolution. (This can be immediately shown when
$\Gamma(E) $ can be approximated by an energy independent
operator). Two important points must be emphasized. First, the
discussion presented above is valid only when the bath spectrum is
continuous. Second, $\Gamma(E) $ vanishes unless $E$ is in the
support of the spectral measure. Similar expressions appear e.g.,
for the width of resonances in the Breit-Wigner theory \cite
{BreitWigner}.

 Various approximations and models where used to
extract the main features of quantum irreversibility as described
by (\ref{eq:dissipative}). Amongst the most popular is the
harmonic-bath model where the bath consists of a continuous set of
harmonic oscillators which are linearly coupled to the system
degrees of freedom \cite {Feynman}. In this case, some of the
integrations can be carried out, and using first order
perturbation theory, the decay width introduced by the bath can be
computed explicitly.

The purpose of the present work  is to carry out the above program
for a system which consists of a graph that is coupled linearly to
a harmonic bath. The system will be described in section
(\ref{sec:system}) and the corresponding Schr\"odinger operator
will be discussed. Because of the form of the coupling, the
question of finding a self-adjoint extension is not simple. In the
present work, a physically motivated form of the Schr\"odinger
operator  is constructed, and the rigorous proof of its validity
is given in an adjacent paper by Michael Solomyak
\cite{michaelnext}.

 Irreversibility is demonstrated by computing the vertex scattering
matrices \cite {KS97,KS99}, which are responsible for the transfer
of current between the bonds of the graph (section
({\ref{sec:vertex})). For sufficiently strong coupling, the vertex
scattering matrices are sub-unitary, which means that the total
probability current is not conserved. Rather, it decreases upon
scattering at the vertex. In other words, the quantum mechanical
evolution is not unitary. The surprise is that this happens even
when a {\em single} harmonic oscillator is coupled at the vertex!
This mechanism should not be confused with what happens naturally
when an additional  lead to infinity is coupled to the vertex, and
its effect on the original scattering is to draw current. This
mechanism which induces an  ``escape width" to the graph dynamics
is not what is considered here. Rather, non unitarity is due to
the coupling to a harmonic oscillator.

The rest of the paper attempts to explain this result in physical
terms. To this end, a simple graph is presented and solved
approximately in section (\ref {sec:model}). Using this model, one
can demonstrate and explain how  the coupling of a single harmonic
oscillator to a compact graph can modify the spectrum in a major
way. If the coupling is week, the spectrum remains pure point.
Beyond a critical strength, a continuous component appears in the
spectrum. The appearance of the continuum occurs at the same value
of the coupling constant at which the vertex scattering matrix
becomes non unitary. A rigorous treatment of this spectral problem
is also provided in  Solomyak's paper \cite{michaelnext}.

\section{The Schr\"odinger operator on a graph and the coupling to a harmonic bath}
\label{sec:system}

A graph ${\cal G}$ consists of $V$ vertices connected by $B$ bonds
according to the connectivity matrix $C_{i,j}$ which takes the
value $1$  if the vertices $i,j$ are connected, and it vanishes
otherwise. We assign the natural metric to the bonds, and
associate a length $L_b$ to each. The position $x$ of a point on
the graph is determined by specifying on which bond $b$ it is, and
its distance $x_b$ from the vertex with the \emph{smaller} index,
$0\le x_b\le L_b$. Sometimes, it will be convenient to denote the
bond connecting the vertices $i$ and $j$ by $(i,j)$.

In the absence of a harmonic bath, the Schr\"odinger operator can
be defined in the following way:  Let $x\in {\mathcal{G}}$ and
$\Psi(x)$ a real valued and continuous function on
${\mathcal{G}}$, so that $\Psi(x)= \psi_b(x_b)$ for $x \in b$, and
$0 \le x_b \le L_b$, where $\psi_b(x_b)$ are twice differentiable
in the interior of the bond. We restrict $\Psi(x)$ to be square
integrable, and require that it is uniquely defined at the
vertices. That is, all the functions $\psi_b$ which correspond to
bonds connected  at a vertex  $i$ attain  the same value $\phi_i$
at the common vertex. Given an arbitrary set of $V$ non-negative
constants $\lambda_i$ one  constructs the quadratic form:
\begin{equation}
\label{eq:quad0} L_0[\Psi] = \sum_{b=1}^B   \int_0^{L_b} {\rm
d}x_b \left| \frac{{\rm d} \psi_b(x_b)}{{\rm d}x_b} \right |^2 +
\sum_{i=1}^V \lambda_i\phi_i^2 \ .
\end{equation}
The Euler-Lagrange variational principle selects the stationary
solutions of the quadratic form, as the solution of the
Schr\"odinger equation on the bonds:
\begin{equation}
\label{eq:schroedinger0}
\left (-\frac   {{\rm d^2 \ }}{{\rm d} x_b^2} -E\right ) \psi_b(x_b) =0 \ ,
\end{equation}
subject to  the boundary conditions
\begin{equation}
\label{eq:boundary0}
 \sum_{j=1}^V\left .C_{i,j}\frac{{\rm d \ \ \ \
}}{{\rm d} x_{(i,j)} }\ \psi_{(i,j)}(x_{(i,j)})\right
|_{i}-\lambda_i\phi_i =0 \ .
\end{equation}
The summation is over all the bonds $(i,j)$  which emanate from
the vertex $i$, and the derivatives are computed at the vertex.
The quadratic form (\ref {eq:quad0}) is positive definite, and
therefore, the boundary condition  (\ref{eq:boundary0}) provides a
self adjoint extension of the Schr\"odinger operator for any
choice of the (non-negative) constants $\lambda_i$. This operator
determines the quantum dynamics of the ``system". One can extend
this operator to non compact graphs, by adding bonds from the
vertices to infinity. The corresponding scattering problem was
amply discussed in the literature
\cite{KS99,KoSch99,KS00,Kuchment02,KS03}.

The bath degrees of freedom consist of $M$ harmonic oscillators
with coordinates $q_m$, and frequencies $\omega_m$, subject to the
hamiltonian
\begin{equation}
\label{eq:hoham} h_m= \frac{1}{2}\left( -\frac{{\rm d^2  \ }}{{\rm
d} q_m^2 } + \omega_m q_m^2\right) \ ;\ q_ m\in \mathbb{R} \ .
\end{equation}
Denote ${\bf q} =(q_1,\cdots,q_M)$ and $H_{osc}= \sum_{m=1}^M
h_m$. $H_{osc}$ is the ``bath" hamiltonian. The coupling of the
bath to the system is introduced by extending the quadratic form
(\ref {eq:quad0}) to include the graph and harmonic bath
coordinates. Consider $\Psi(x,{\bf q}) = \psi_b(x_b,{\bf q})\ {\rm
for} \ x \in b $, where each of the $\psi_b(x_b,{\bf q})$ is
square integrable. For every value of ${\bf q}$, $\Psi(x,{\bf q})$
is uniquely defined at the vertices, where, at the vertex $i$ it
assumes the value $\phi_i({\bf q})$. To each vertex, we assign a
real valued function $\Lambda_i({\bf q})$, which is bounded at any
finite domain of $\mathbb{R}^M$ and is allowed to diverge
algebraically as $|{\bf q}|\rightarrow \infty$. (In the following,
we shall refer to these functions as the ``form-factors" ). We
construct the quadratic form,
\begin{eqnarray}
\label{eq:quadho}
 \hspace{-10mm}
L_{osc}[\Psi] &=& \sum_{b=1}^B \int {\rm d}^M q \left \{ \right.
\int_0^{L_b} {\rm d}x_b  (\ \left| \frac{{\rm d} \psi_b(x_b,{\bf
q})}{{\rm d}x_b} \right |^2 +\psi^*_b(x_b,{\bf q}) H_{osc} \psi
_b(x_b,{\bf q}\ )
  ) \nonumber \\
&+&\sum_{i=1}^V \Lambda_i({\bf q})\phi_i^2({\bf q})\left. \right
\} \ .
\end{eqnarray}
Again, the Euler-Lagrange variational principle selects the
stationary solutions of the quadratic form, as the solution of the
Schr\"odinger equation on the bonds:
\begin{equation}
\label{eq:schroedingerh}
\left (-\frac   {{\rm d^2 \ }}{{\rm d} x_b^2}+H_{osc} -E\right ) \psi_b(x_b,{\bf q}) =0 \
 ,
\end{equation}
subject to  the boundary conditions
\begin{equation}
\label{eq:boundaryh} \left . \sum_{j=1}^V  C_{i,j} \frac{{\rm d \
\ \ }}{{\rm d} x_{(i,j)} }\ \psi_{(i,j)}(x_{(i,j)},{\bf q})\right
|_{i}-\Lambda_i({\bf q})\phi_i({\bf q})=0 \ .
\end{equation}
The coupling between the ``system" and the harmonic ``bath" is
thus mediated through the boundary conditions (\ref
{eq:boundaryh}).

When the form-factors $\Lambda_v({\bf q})$ are non negative the
quadratic form $L_{osc}[\Psi]$ is positive definite, and the
Schr\"odinger operator (\ref {eq:schroedingerh}), subject to the
boundary conditions (\ref {eq:boundaryh}) is self adjoint. The
standard {\it linear} coupling of harmonic baths  is written as
\begin{equation}
\label{eq:lincoup} \Lambda_i({\bf q}) = \sum_{m=1}^M \Lambda_{i
m}\ q_m \ ,
\end{equation}
with $\Lambda$ a  $V \times M$ matrix with ${\bf q}$-independent
real coefficients. Thus, $L_{osc}[\Psi]$ is not definite, and  the
proof that the system (\ref{eq:schroedingerh},\ref{eq:boundaryh})
is self-adjoint requires a special treatment, which is provided in
\cite{michaelnext}. The intuitive motivation behind this coupling
scheme is that for a vertex with $v=2$, and a single harmonic
oscillator, the boundary condition at the vertex is equivalent to
an effective potential  $\Lambda q \delta(x)$. This represents a
well localized coupling of the particle on the graph to the
harmonic oscillator.

\section{The vertex scattering matrix }
\label{sec:vertex}

The boundary conditions at the vertices of graphs (without
coupling to any harmonic oscillators) (\ref{eq:boundary0}) impose
the conservation of the probability current across the vertex. In
other words, the boundary conditions at a vertex can be described
 by a unitary \textit{vertex scattering matrix} which provides
the linear relation between the amplitudes of the waves which
imping upon the vertex and the waves scattered from it
\cite{KS97}. The vertex scattering matrix which corresponds to the
boundary conditions (\ref{eq:boundary0}) can be written explicitly
as
\begin{equation}
\sigma^{(i)}_{b,b'}  (k)\ =\  -\delta_{b,b'} + \frac{ 1+{\rm
e}^{-2 i \arctan \frac{\lambda_i}{v_i k}} }{v_i}
 \label{eq:vscat0}
\end{equation}
where $i$ denotes the vertex under consideration, $b$ and $b'$
denote the bonds which emanate from the vertex, and $k={\sqrt E}$
is the wave number under consideration. Note that this  $v_i\times
v_i$ vertex matrix distinguishes only between reflection ($b= b'$)
and transmission between different bonds ($b\ne b'$), which is
independent of their specific identity. This is an expression of
the fact that the boundary condition at the vertex is invariant
under the interchange of the bonds. $\sigma^{(i)}_{b,b'} (k)$ is
symmetric, and its  unitarity can be easily demonstrated. The
single vertex, and the bonds emanating from it, is some times
referred to as a ``star-graph".

Describing the quantum evolution and dynamics in terms of waves
travelling freely along the bonds  and scattered at the vertices,
is a very useful concept in the theory of quantum graphs
\cite{KS97}. This is also true when harmonic oscillators are
coupled, and we shall dedicate the rest of this section to the
derivation of the the vertex scattering matrices which correspond
to the boundary conditions (\ref{eq:boundaryh}). We shall show,
that the coupling of the harmonic oscillators can have a profound
effect - the vertex scattering matrix is not necessarily unitary
for large values of the coupling $\Lambda$. Thus, the evolution on
the graph is not conserving and hence irreversible!

To simplify the presentation,  we consider the coupling to a {\it
single} harmonic oscillator ($M=1$), and choose the units such
that $\omega_1=1$. The spectrum of the combined system of a
harmonic oscillator and the star graph without coupling, consists
of overlapping continua with thresholds at $E_n=n+\frac{1}{2}$.
Further simplification is achieved by considering energies in the
interval  $\frac{1}{2} < E < E_1$. In this energy range, and for
$x$ values away from the vertex, only the ground state ($n=0$) of
the harmonic oscillator is populated, which corresponds to purely
elastic scattering. We shall compute the vertex scattering matrix
by considering an incoming wave incident on the bond $b'$ only.

Denoting
\begin{equation}
 k^2 = E-\frac{1}{2} \ \ ;\ \   k_0=k\ \ ; \ \
 k_n=\sqrt{n-k^2}\ \  {\rm for}\ \  n\ge 1 ,
\end{equation}
 we expand the wave function in
the complete orthonormal basis $\chi_n(q)$ of eigenfunctions of
the harmonic oscillator hamiltonian $h_1(q)$. The bond wave
functions are defined on the positive half line, with the common
vertex at  $x_b = 0$. The expansion takes the form
\begin{equation}
\label{eq:expand1} \psi_b(q,x_b)=\frac{1}{\sqrt{k_0 }}\
\eta^{(b)}_0(x_b) \ \chi_0(q)\ +\ \sum_{n=1}^{\infty}
\frac{a^{(b)}_n}{\sqrt{k_n }}\ \eta^{(b)}_n(x_b)\ \chi_n(q) \ .
\end{equation}
The functions $\eta^{(b)}_n(x_b)$ are given  by
 \begin{eqnarray}
 \label{eq:w.f.1}
\eta^{(b)}_0(x_b)&=&\left\{
\begin{array}{l}
  {\rm e}^{-ik_0 x_b} +r\ {\rm e}^ { ik_0 x_b} \ \ \ \ \ b=b'  \\  \\
 \ \ \ \ \ \ \ \ \ \ \ \ t\ {\rm e}^ {\ ik_0 x_b} \ \ \ \ \ b \ne b'
\end{array}
\right .   \\
\eta^{(b)}_n(x_b)&=&\ \ \ {\rm e}^{-k_n x_b}
\ \ \ \ \ \ {\rm for} \ \ \ \ \ \ \   n\ge 1 \ . \nonumber
 \end{eqnarray}
This way, $\Psi(x,q)$ corresponds asymptotically to an incoming
wave together with scattered waves in the ground state channel. In
all other (closed) channels, the wave functions  are evanescent
along the bonds. Note that in (\ref{eq:w.f.1}) we made use of the
expected invariance of the scattering matrix under the interchange
of bonds, so that $r$ stands for the reflection (diagonal)
elements of the $\sigma$ matrix, and $t$ stands for the
transmission (off-diagonal) elements of $\sigma$. Both parameters
are independent of $b$.

The continuity at of $\Psi(x,q)$ at the vertex implies that the
expansion coefficients ${a^{(b)}_n}$ are the same for all $b$, and
therefore they will be denoted in what follows by $a_n$. Moreover,
\begin{equation}
1+r=t \label{eq:bc}
\end{equation}
Substituting in  (\ref {eq:boundaryh}) and using (\ref{eq:bc}), we
get
\begin{equation}
 t\ = \frac{2}{v} \ +\  \frac{\lambda}{2 i} \frac{1}{\sqrt{k_0k_1}}\ a_1 ,
 \label{eq:bc1}
\end{equation}
and,
\begin{eqnarray}
\label{eq:scat1}
 2 a_1+\lambda\left (\  \ a_2\ \ \sqrt{\
\frac{2}{k_1 \ k_2}} \ \right )
 &=&- \frac{\lambda}{\sqrt {k_0\ k_1}}\ t  \\
 2 a_n +\lambda \left( a_{n+1}\sqrt{\frac {n+1}{k_nk_{n+1}}}+
a_{n-1}\ \sqrt{ \frac {n }{k_n k_{n-1}}}\ \right) &=&\ \ \ \ \ \ \
0 \ \ \ \ \ \ \ {\rm for}\  \  n\ge 2 \ .  \nonumber
\end{eqnarray}
Where, $\lambda\equiv \frac{{\sqrt 2}\Lambda}{v} $.
 (\ref {eq:scat1}) is an infinite set of linear, inhomogeneous equations, which
are defined in terms of the infinite Jacobian matrix
$J(k,\lambda)$ whose elements are explicitly given in
(\ref{eq:scat1}). For large $n$
\begin{equation}
\sqrt{\frac{n}{k_nk_{n-1}}} = 1 + {\cal O}\left( \frac{1}{n}\right
 )\  .
\label{eq:coefestimate}
\end{equation}
We shall denote by $J_0(\lambda)$ the Jacobian matrix whose
elements are defined by the equations with constant coefficients
 \begin{equation}
\label{eq:iterasym}
 2 a_n + \lambda \left( a_{n+1} + a_{n-1} \right) =0 \
\end{equation}
which approximate (\ref{eq:scat1}).

Denoting by $G(z;k,\lambda)$ the resolvent
\begin {equation}
G(z;k,\lambda) = \frac{1}{J(k,\lambda)-z} \label{eq:resolvent}
\end{equation}
we get
\begin{equation}
a_1= - \lim_{\epsilon \nearrow 0}
G_{1,1}(i\epsilon;k,\lambda)\frac{\lambda}{\sqrt {k_0\ k_1}}\ t .
\label{eq:a1}
\end{equation}
Denote
\begin{equation}
\gamma(k,\lambda)=\frac{\lambda^2}{2\ k_0\ k_1} \lim_{\epsilon
\nearrow 0} G_{1,1}(i\epsilon;k,\lambda) \ .
 \label{eq:gamma1}
\end{equation}
 Substituting (\ref{eq:a1}) in (\ref{eq:bc1}), we find
\begin{equation}
t\ =\ \frac{\frac{2}{v}}{1-i \gamma(k,\lambda)} \ \ \ \ ; \ \ \ \
\ r\ = \ t-1\ =\frac{(-1+\frac{2}{v})+ i \gamma(k,\lambda)}{1-i
\gamma(k,\lambda)} \ .
  \label{eq:tandr}
\end{equation}
Note that when $\lambda=0$, the reflection and transmission
coefficients reduce to (\ref {eq:vscat0}) (with $\lambda_i=0$).
The flux in the leads is conserved if $(v-1) |t|^2+|r|^2=1$, which
is satisfied only if ${\cal I}m (\gamma(k,\lambda)) =  0. $ We
shall now show that this happens if and only if $|\lambda| < 1 $.
In other words, the dissipation of flux, or equivalently, quantum
irreversibility occurs for $|\lambda| > 1 $.

Following \cite{kil-simon}, we associate to the Jacobi matrix $J$
the spectral measure $\mu(x;k,\lambda)$, defined by
\begin{equation}
G_{1,1}(z;k,\lambda)=\int\frac{{\rm d} \mu(x;k,\lambda)}{x-z}
\end{equation}
Using the estimate (\ref{eq:coefestimate}), we find that the
operator $(J-J_0)$ is Hilbert-Schmidt. The main theorem in
\cite{kil-simon} guarantees that the absolutely continuous
component of $\mu(x;k,\lambda)$ is supported in
$\sigma(k,\lambda))=[2(1-|\lambda)|,2(1+|\lambda)|]$.

As long as $|\lambda|<1$, the support $\sigma(k,\lambda))$ is on
the positive half-line. The limit $z=i \varepsilon \nearrow 0$ is
straight forward, resulting in a real value for
$\gamma(k,\lambda)$. Hence for $|\lambda| < 1 $ the flux in the
$x$ channel is conserved.

When $|\lambda|>1$, the interval $\sigma(k,\lambda))$ includes the
value $0$. To compute (\ref{eq:gamma1})  we use
(\ref{eq:deltaid}), and get
\begin{equation}
{\cal I}m\gamma(k,\lambda)= -\frac{\pi\lambda^2}{2
k_0k_1}\varrho(0,\lambda) \label{eq:gamma2} \ne 0
\end{equation}
Where we write  ${\rm d} \mu(x;k,\lambda)=\varrho(x,\lambda){\rm
d}x$. Thus, conservation of flux is violated, which is equivalent
to the onset of irreversibility. This is the main result of the
present work. It proves the fact that ${\cal
I}m\gamma(k,\lambda)\ne 0$, but it does not provide an explicit
expression for $\gamma(k,\lambda)$. However, once it is computed
numerically or approximately (see next paragraph) it can be used
together with (\ref {eq:tandr}) to compute the vertex scattering
matrix.
\begin{equation}
\sigma_{b,b'}(k,\lambda) = -\delta_{b,b'} + t(k,\lambda)
 \label{eq:sigkg}
\end{equation}
An approximate computation of the resolvent for energies in the
vicinity of the elastic threshold ($E \gtrsim\frac{1}{2}$) can be
obtained by replacing the Jacobi matrix $J(k,\lambda)$ by
$J_0(\lambda)$. It is has constant values along its diagonals, and
an exact solution of the inhomogeneous equations for $\lambda>1$
read
\begin{equation}
 a_n =   {\rm e}^{in\alpha} \frac{
 t}{\sqrt{k_0k_1}} \ \ ; \ \ \cos \alpha = -\frac{1}{\lambda}
 \label{eq:exactGJ0}
\end{equation}
Thus,
\begin{equation}
 \gamma(k,\lambda) \approxeq - \frac{\lambda }{k_0k_1} \
{\rm e}^{ i \alpha} \ .
\end{equation}
 This is an explicit expression which can be directly used in(\ref
 {eq:sigkg}).

Note that the valency $v$ of the vertex enters only though the
scaled strength parameter $\lambda= \frac {\sqrt{2} \Lambda}{v}$.

 When we relax the condition $E<E_1$ the same phenomenon will
occur albeit it involves now the inelastic scattering channels,
whose number is finite and determined by the total energy $E$. It
gives no new insight, and therefore it will not be discussed
further.

The fact that the spectrum of the Jacobi matrix is continuous
played a crucial role in the previous arguments. It appears in a
way which is reminiscent of the discussion of quantum
irreversibility presented in the introduction.

 Coming back to the observation that the ingoing current
exceeds the total outgoing current, one naturally asks where can
one find the rest of the probability density. This, and other
relevant questions will be clarified in the next section.

\section {A simple model}
\label{sec:model}

The star graph with $v=2$ is equivalent to the Schr\"odinger
equation

\begin{equation}
\label{eq:model}
 \left [  -\frac{ \partial ^2  \  }{\partial x^2 }
+ \frac{1}{2} \left( -\frac{\partial^2 }{\partial q^2 } +
q^2\right) + \Lambda\ q\  \delta(x) \right ]\Psi(x,q) =E\
\Psi(x,q) \ ,
\end{equation}
with $(x,q) \in \mathbb{R}^2$. The boundary condition at the
vertex is replaced by a term which can be interpreted as a
potential, linear in $q$ and well localized in $x$. The
Schr\"odinger equation describes a particle in the $(x,q)$ plane,
moving under the action of the potential shown in figure (\ref
{fig:pot}), where the $\delta$ function is replaced by a narrow
Gaussian.
\begin{figure}[h]
\hspace*{15mm}
 \epsfxsize.6\textwidth
  \epsfbox{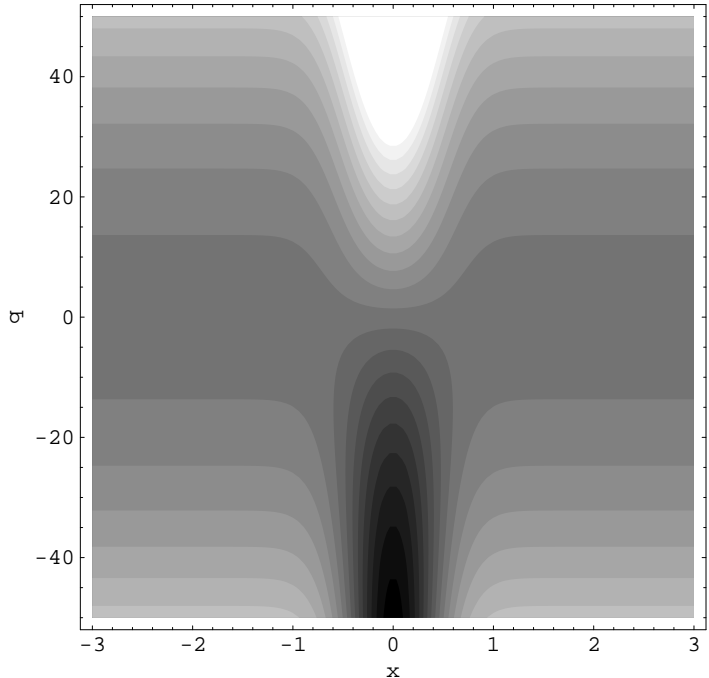}
\caption {Grey level map of the potential in (\ref {eq:model})}
\label{fig:pot}
\end{figure}
In this figure, the hight of the potential is indicated by the
grey level, where black (white)  marks the domains where the
potential is most negative (positive). Consider now a particle
moving in  two dimensions and acted upon by this potential.  Away
from the domain of strong coupling, the potential is independent
of $x$ and quadratic in $q$, thus forming a valley along the $x$
axis. Asymptotically, the particle moves with a constant speed in
the $x$ direction, while oscillating in the $q$ direction. At the
vicinity of $x=0$, the potential changes abruptly. Here a positive
ridge (assuming $\lambda >0$) along the positive $q$ axis makes
the $q>0$ domain inaccessible to the particle, while a down
sloping valley pointing towards the negative $q$ axis, may attract
the particle. As long as we use a finite Gaussian to simulate the
$\delta$ potential, the quadratic potential in the  $q$ direction
will take over, and will form a barrier from which the particle
will be reflected back and eventually the particle will find
itself moving along the $x$ axis in either the positive or the
negative directions. The limit where the Gaussian becomes a
$\delta$ function, is delicate, and our computations show that it
depends on the size of $\lambda$. Once $|\lambda|>1$, the
effective potential valley extends to infinity, and it is
sufficient to  to draw flux in the $q$ direction without ever
reflecting it back. Going back to the original way of looking at
the problem, we can say that, with a finite probability,  the
``system" degree of freedom remains near the vertex at  $x=0$
while the harmonic oscillator is excited in an outgoing stretching
mode due to the strong interaction.

Another point of view can be obtained by studying the stationary
wave function $\Psi(q,x)$ (\ref{eq:expand1})  at $x=0$. We
consider  its projection on the space of functions spanned by the
oscillator states with $n\ge1$, which correspond to evanescent
modes in the $x$ direction,
\begin{equation}
\label{eq:expandx=0}
 \Psi(q,0)=\ \sum_{n=1}^{\infty} \frac{a_n}{\sqrt{k_n }}\ \chi_n(q)
 \ .
\end{equation}
 We  use the approximate
expression (\ref{eq:exactGJ0}) for $a_n$, and  replace the
harmonic oscillator wave functions by their WKB approximation,
\begin{eqnarray}
  \chi_n(q) &\approx& \sqrt{\frac{2}{\pi}}\frac{\cos
S(n)}{(2n+1-q^2)^{1/4}} \ , \nonumber \\
 S(n)&=& (n+\frac{1}{2}) \left(
\arcsin\frac{q}{\sqrt{2n+1}} +\frac{q}{\sqrt{2n+1}} (1-\frac{q^2}{
2n+1 })^{\frac{1}{2}} -\frac{\pi}{2}
 \right)\ .
\end{eqnarray}
Substituting in (\ref{eq:expandx=0}), and using the saddle point
approximation to perform the sum, we obtain:
\begin{equation}
 \Psi(x=0,q) \approx C\cdot \frac{  {\rm e}^{-i
 \frac{q^2}{2} \tan  \alpha  }}
 {\sqrt {2\pi i \cos  \alpha  }}
\end{equation}
Where $C$ is independent of $q$.  This function is  {\it not}
square normalizable.

The simple computation above provides some heuristic evidence in
support of the suggestion that the interacting system possesses a
new continuum channel, where the particle can be trapped
indefinitely. This issue is discussed in great detail and rigor in
Solomyak's paper \cite {michaelnext}.  In the sequel we shall
present another variant of the simple model, where the onset of a
continuum at $|\lambda| >  1$ will be elucidated from yet another
perspective.

Consider (\ref {eq:model}) subject to the boundary conditions
\begin{equation}
\label{eq:modbound}
\Psi(x=-L_1,q)=\Psi(x=L_2,q)=0 \ .
\end{equation}
This  Schr\"odinger operator describes a particle which is
confined to  the interval $[-L_1,L_2]$ and a harmonic oscillator,
which are coupled by an interaction which is very well localized
at $x=0$, and is linear in the oscillator amplitude. The spectrum
of the uncoupled ($\Lambda=0$) problem is pure point:
\begin{equation}
\label{eq:spect} E_{n,m} = (n+\frac{1}{2}) + \left (
\frac{\pi}{L_1+L_2}\right)^2  m^2 \ \ ;\ \  n=0,1\cdots\ ; \
m=1,2,\cdots
\end{equation}

We shall turn now to the case $\Lambda \ne 0$. We consider in
particular the domain of energies $E<\frac{1}{2}$, just under the
lowest energy of the unperturbed system.   Again we expand
$\Psi(x,q)$ in the complete orthonormal basis $\chi_n(q)$, and use
the notation $k_n=+\sqrt{ \frac{1}{2}+n -E} $ for $n=0,1,\cdots$.
\begin{equation}
\label{eq:expand}
\Psi(q,x)=\sum_{n=0}^{\infty} \frac{a_n}{\sqrt{k_n }}\ \eta_n(x)\ \chi_n(q) \ ,
\end{equation}
where $\eta_n(x)$  satisfy the boundary conditions
(\ref{eq:modbound}), are continuous at the origin, and normalized
by   $\eta_n(0)=1$. Explicitly,
 \begin{eqnarray}
\eta_n(x)&=&\left\{
\begin{array}{l}
\frac{\sinh k_n(L_1+x)}{\sinh k_n L_1} \ \ \ \ \ x \le 0  \\  \\
\frac{\sinh k_n(L_2-x)}{\sinh k_n L_2} \ \ \ \ \ x \ge 0
\end{array}
\right . \ \ \ \ \ \ \   \nonumber \label{eq:w.f.}
\end{eqnarray}
These functions are evanescent away from the origin  $x=0$.
Substituting (\ref{eq:expand}) in (\ref {eq:model}), we get a set
of equations from which the coefficients $a_n$ are to be
determined.
\begin{equation}
\label{eq:iter}
 \gamma_n a_n +\lambda \left( a_{n+1}\sqrt{\frac {n+1}{k_nk_{n+1}}}+
a_{n-1}\sqrt{\frac {n }{k_n k_{n-1}}}\right) =0 \ ,
\end{equation}
where, $\lambda = \frac{\Lambda}{\sqrt{2}}$ and
 \begin{equation}
\gamma_n\  =
 \coth k_nL_1+\coth k_n L_2  \ \ \  .
\end{equation}

Equation (\ref {eq:iter}) is a second order recursion relation
with the boundary condition $a_{-1}=0$. The coefficients depend on
the energy E, and the spectrum of the Schr\"odinger operator is
determined as the values of $E$ for which the series
$a_n(E)/\sqrt{k_n}$ is square summable.  Before we proceed to
analyze this further, we should point out that  in the limit $n
>> E $,
\begin{equation}
\sqrt{\frac{n}{k_nk_{n-1}}} = 1 + {\cal O}\left( \frac{1}{n}\right
) \ \ \  ; \ \ \  \gamma_n  =1 + {\cal O}\left( \frac{1}{n}\right
) \ . \label{eq:coefestimate1}
\end{equation}
Thus, (\ref {eq:iter}) limits  to the recursion relation with
constant coefficients (\ref{eq:iterasym}). This is of crucial
importance to our discussion.

We shall now provide the heuristic  arguments which show that for
$|\lambda|<1$ the spectrum of the Schr\"odinger operator (\ref
{eq:model}) retains its pure point character. However, for
$|\lambda|>1$ it has a purely continuous component.  Using the
initial conditions $a_{-1}=0,\ a_0=1$, one can apply the exact
recurrence relation (\ref {eq:iter}) and obtain $a_{N}(E)$ for
arbitrary large  $N$. Chose $N$ such that for $n \ge N$ one
introduces only a small error if one replaces (\ref {eq:iter}) by
(\ref {eq:iterasym}). The solution of (\ref {eq:iterasym}) can be
written explicitly:
\begin{equation}
\label{eq:approxan}
a^{approx}_n = A\ \xi ^{ n}  + B\ \xi ^{- n} \ \ \ ;\ \ \ \xi=
\frac{1}{\lambda}\left(-1+\sqrt{1-\lambda^2}\right ) \ ,
\end{equation}
where $A,B$ are arbitrary constants, to be determined by the initial conditions.

For $|\lambda| <1$,\ $|\xi|<1$. Thus, to get a converging solution $B$ must vanish. The
matching of the approximate series at $n=N$ requires
\begin{equation}
\label{eq:secapprox}
\xi = \frac{a_{N+1}(E)} {a_{N}(E)} \ ,
\end{equation}
which is the spectral secular equation. Limiting our attention to the spectral
interval $E < \frac{1}{2}$, and with $L_1,L_2$ sufficiently large, we can chose $N=0$, and
the secular equation reads,
\begin{equation}
\label{eq:secapprox0}
\xi = \frac{a_{1}(E)} {a_{0}(E)} \  = -\frac{2}{\lambda}\left
((\frac{1}{2}-E)(\frac{3}{2}-E))\right )^{\frac{1}{4}}.
\end{equation}
Thus,
\begin{equation}
E =1-\frac{1}{2}\sqrt{1+ \frac{1}{4}\left(1-\sqrt{1-\lambda^2}\right )^4}\ \ <\ \frac{1}{2}
\ .
\end{equation}
The approximate solution predicts the existence of a bound state
of the system below the ground state of the unperturbed harmonic
oscillator. This is rigorously proved in \cite{michaelnext}.

For $|\lambda| >1$,\  $\xi=\exp(-i\arctan (\sqrt{\lambda^2-1}) =
{\rm e}^{-i\zeta}$. Now, both terms in (\ref{eq:approxan})
contribute, and the secular equation reads,
\begin{equation}
\label{eq:secapprox1}
 \frac{a_{N+1}(E)} {a_{N}(E)}= \frac { A\ {\rm e}^{-i(N+1)\zeta}  + B\
{\rm e}^{ i(N+1)\zeta}} { A\ {\rm e}^{-iN \zeta}  + B\ {\rm e}^{iN\zeta}}\ .
\end{equation}
In contrast with the previous case, for every $E<\frac{1}{2}$,
 the matching condition are satisfied when
\begin{equation}
\frac{A}{B} = -{\rm e}^{2i(N+1)\zeta}\ \frac{1-\frac{a_{N+1}(E)} {a_{N}(E)}\ {\rm
e}^{-i\zeta}} {1-\frac{a_{N+1}(E)} {a_{N}(E)}\ {\rm e}^{\ i\zeta}} \ .
\end{equation}
Hence, the secular equation can be solved for every $E$, and the
spectrum is continuous.  Moreover, the $a_n(E)$ do not vanish in
the limit of large $n$. Substituting the $a_n$ in (\ref
{eq:expand}), the resulting $\Psi(x,q)$ is bounded but not square
integrable, which is typical for  eigenfunctions in the continuous
spectrum. We emphasize again the fact that the onset of
irreversibility coincides with the appearance of a continuum. This
result, which was illustrated above for the simple model, is
general, and is entirely due to the typical form of the boundary
conditions at the vertices.

A last comment is in order. Often, when facing an infinite set of
equations of the type studied here, it is tempting to truncate
them at a physically or computationally motivated point. In the
present context, one would naturally attempt to truncate the space
of closed channels, and perform the computations within the
subspace of opened channel, exclusively. This would {\it always}
result in a unitary description of the vertex scattering matrix,
and the possibility of flux dissipation would be overlooked.

To summarize - the main lesson from this work is  that genuine
quantum flux dissipation and irreversibility  can be induced by a
single harmonic oscillator - the transmitted and the reflected
fluxes {\it do not} add up to the incoming flux. This result is of
general interest because it shows that in a strongly interacting
system, continuum channels, which are absent in the uncoupled
system and bath, can be opened by the interaction, and induce
dissipation. At such instances, the perturbative approach to
dissipation might not reflect the true nature of the problem.

 \section{Acknowledgements}
 \label{sec:acknowledgement}
First and foremost, the author wishes to thank  Professor Michael
Solomyak for the enjoyable discussions and exchanges which
accompanied the work reported in the present paper. Professor
Solomyak's critical and pedagogical comments were most valuable.
Part of this research was conducted when the author was a guest at
the Mittag-Leffler Institute (MLI) and the Technical University
(KTH) in Stockholm. Many thanks are due to Professor Ari Laptev
and the MLI and KTH stuffs for their kind hospitality. Support
from the Minerva Center for Nonlinear Physics, the Israel Science
Foundation and the Minerva Foundation are acknowledged.

\end{document}